# Energy Level Engineering in $In_xGa_{1-x}As/GaAs$ Quantum Dots Applicable to QD-Lasers by Changing the Stoichiometric Percentage

**Mahdi Ahmadi Borji[*] and Esfandiar Rajaei[**]**
**Department of Physics, the University of Guilan, Rasht, Iran**
*Email: Mehdi.p83@gmail.com, **Corresponding author, E-mail: Raf404@guilan.ac.ir

**ABSTRACT**

Band edge and energy levels of truncated pyramidal $In_xGa_{1-x}As$/GaAs (001) quantum dots are studied by single-band effective mass approach, and the dependence to stoichiometric percentages is investigated. It is shown that enhancement of indium percentage decreases the band gap and the recombination energy of electrons and holes. Our principal result is that decrease of recombination energy and band gap is nonlinear and the slopes are different band gap and e-h recombination energy. In addition, it is proved that strain tensor is diagonal along z-axis and the absolute value of the components gets larger by more indium inclusion. Our results appear to be in very good consonance with similar studies.

## I. INTRODUCTION

Engineering of nanostructures enables to control their strain effects, energy gap, energy levels, band structures, etc. Optimization of these parameters gives rise in the optimized performance of photonic devices such as many types of lasers. Semiconductor nanostructures include quantum wells, wires, and dots in which carriers are respectively confined to one, two, or three dimensions. Quantum confined semiconductors have been the focus of many studies because of their optical properties arising from quantum confinement of electrons and holes [1-5]. In semiconductor quantum dots (QDs), carriers are limited to a three-dimensional box smaller than 100nm. The confinement arises when the band gap of the QD nanomaterial is less than that of the surrounding semiconductor [6]. Progress in fabrication of QD-lasers has recently attracted a wide discussion on the application of quantum systems in optoelectronics [7-10].

Many research groups attempt to optimize QD lasers [7, 11-13]. InAs based QDLs are the most interesting cases in the local area networks because of the low threshold current, enhanced modulation speed, and temperature stability [14]. QDs can be optimized by changing the compositions; many scientists are interested in $In_xGa_{1-x}As/GaAs$ QD lasers due to their interesting and applicable features [4, 11, 12, 15, 16]. Despite the wide range of researches, there is in the literature a lack of theoretical studies elaborating to find composition effects in QDs. Change of indium percentage can lead to formation of QD with different sizes [17].

Self-assembled quantum dot structures are formed from in the epitaxial Metal Organic

Chemical Vapor Deposition (MOCVD) or Molecular Beam Epitaxy (MBE) [18, 19]. Stranski-Krastanov mode concerns with strained heteroepitaxial growth, which starts by layer-by-layer growth but after deposition of a number of layers suddenly a transition occurs to formation of islands. If the lattice constant of deposited atoms is greater than the ones for substrate by just a little difference, quantum dot islands are formed with a few nanometers in size [20]. The shape and size of the QDs is connected to the epitaxy conditions and type of heterostructures. QD formation can be controlled by factors dealing with free energy minimization of crystals. So, the distribution and sizes are random which leads to different characteristics of the output laser properties [21]. It is shown that change of stoichiometric percentage in epitaxial growth of QDs influences on both QD size and density, and changes the electron-hole recombination energies [22-24]. Some theoretical and experimental works elaborated to find the relationship between QD size and the mismatch of the grown QD with the substrate [17, 25]. However, growth time [25] and temperature [26] can change the condition and so it is possible to have QDs of different indium ratios, but with the same sizes. While changing the indium percentage, we assume the QD size to remain fixed. As the mismatch between lattice constants of substrate and QDs is dependent to substrate index [27], we supposed the QDs to grow on (001) substrate index.

The rest of this paper is organized as follows: section II is devoted to the model and method of the numerical simulation; results and discussions on the indium percentage effects are presented in section III; lastly, we make a conclusion in section IV.

**II. MODEL AND SIMULATION**

Wave function of electron satisfies the Schrödinger equation

$$H\psi_{nk}(r) = E_n(\mathbf{k})\psi_{nk}(r) \quad (1)$$

In which $\mathbf{k}$ stands for the electron wave vector, $r$ represents the position vector in the lattice, and due the periodicity of lattice, $\psi$ can be written as:

$$\psi_{nk}(r) = e^{ik.r}u_{nk}(r) \quad (2)$$

In which $u_{nk}(r)$ is a periodic Bloch spinor. The Hamiltonian for a strained zinc-blende heterostructure is composed of five terms, i.e., $H = H_0 + H_k + H_{k.p} + H_{s.o.} + H'_{s.o.}$. The used terms read,

$$H_0 = \frac{p^2}{2m_0} + V_0(r, \varepsilon_{ij}) \quad (3)$$

With $V_0(r, \varepsilon_{ij})$ as the periodic potential of the strained crystal substituted instead of $V_0(r)$ when there was no strain; Here $i, j$ stand for $x, y, z$, $m_0$ is the mass of electron, and $\mathbf{p} = -i\hbar\nabla$ is the momentum. The kinetic and $\mathbf{k}.\mathbf{p}$ terms are

$$H_k = \frac{\hbar^2 k^2}{2m_0} \quad (4)$$

$$H_{k.p} = \frac{\hbar}{m_0}\mathbf{k}.\mathbf{p} \quad (5)$$

And the spin-orbit interaction terms are

$$H_{s.o.} = \frac{\hbar}{4m_0^2 c^2}\left(\boldsymbol{\sigma} \times \nabla V_0(r, \varepsilon_{ij})\right).\mathbf{p} \quad (6)$$

$$H'_{s.o.} = \frac{\hbar}{4m_0^2 c^2}\left(\boldsymbol{\sigma} \times \nabla V_0(r, \varepsilon_{ij})\right).\hbar\mathbf{k} \quad (7)$$

With $\boldsymbol{\sigma} = (\begin{bmatrix} 0 & 1 \\ 1 & 0 \end{bmatrix}, \begin{bmatrix} 0 & -i \\ i & 0 \end{bmatrix}, \begin{bmatrix} 1 & 0 \\ 0 & -1 \end{bmatrix})$ as the Pauli spin matrix.

The Schrödinger equation could be solved by expanding $V_0(r, \varepsilon_{ij})$ to first order in strain dyadic, $\varepsilon_{ij}$. This strain can change the basis vectors $\boldsymbol{b}_i$ to $\boldsymbol{b}'_i = \sum_{j=1}^{3} \boldsymbol{b}_j(\varepsilon_{ij} + \delta_{ij})$ and lattice constants $\boldsymbol{a}_i$ to $\boldsymbol{a}'_i = \boldsymbol{a}_i(\sum_{j=1}^{3} \varepsilon_{ij} + 1)$ [28]. By change of variable

$$\boldsymbol{p}' = \boldsymbol{p} + \frac{\hbar}{4m_0 c^2}\left(\boldsymbol{\sigma} \times \nabla V_0(\boldsymbol{r}, \varepsilon_{ij})\right). \quad (8)$$

the equation reads

$$\left(H_0 + \frac{\hbar^2 \boldsymbol{k}^2}{2m_0} + \frac{\hbar}{m_0}\boldsymbol{k}.\boldsymbol{p}'\right) u_{n\boldsymbol{k}}(\boldsymbol{r}) = E_n(\boldsymbol{k}) u_{n\boldsymbol{k}}(\boldsymbol{r}) \quad (9)$$

Which can be solved by the second order non-degenerate perturbation method in which the latter terms are considered as the perturbation. Thus, in the Cartesian space:

$$E_n(\boldsymbol{k}) = E_n(\boldsymbol{0}) + \frac{\hbar^2 k_i k_j}{2}\left(\frac{1}{m_n^*}\right)_{i,j} \quad (10)$$

In which $m^*$ is called the effective mass of electron and can be connected to $m_0$ as [29]:

$$\left(\frac{1}{m_n^*}\right)_{i,j} = \left(\frac{1}{m_0}\right)\delta_{i,j} + \frac{2}{m_0^2}\sum_{m \neq n} \frac{\langle n,0|p'_i|m,0\rangle\langle m,0|p'_j|n,0\rangle}{E_n(0) - E_m(0)} \quad (11)$$

The single-band effective mass approach is used in solving the Schrödinger equation, and the Poisson's equation is solved numerically in a self-consistent manner [30, 31]. All the simulations in this article are conducted under the room temperature, $27\ °C$. Indium percentage of the quantum dot varies from 10% to 100% (i.e., InAs). A quantum dot can be approximated by shapes such as cylinder, pyramid, cube, lens shape [27], …. In our work the QDs are supposed to be truncated pyramids with square base and height of 2/3 times the base side. Moreover, QDs are assumed to be far enough to forbid any influence from other QDs.

Figure 1 shows the profile of a truncated $In_xGa_{1-x}As$ pyramidal QD surrounded by GaAs [32]. Usually, indium percentage of x=0.4 is used in laser devices [33]. Both GaAs and InAs have the zinc blend structures. The pyramid has a square base of area $17nm \times 17nm$ and the height of ~11.3nm on $15nm$ of GaAs and $0.5nm$ of wetting-layer grown on (001) substrate index. The structure is grown towards z direction.

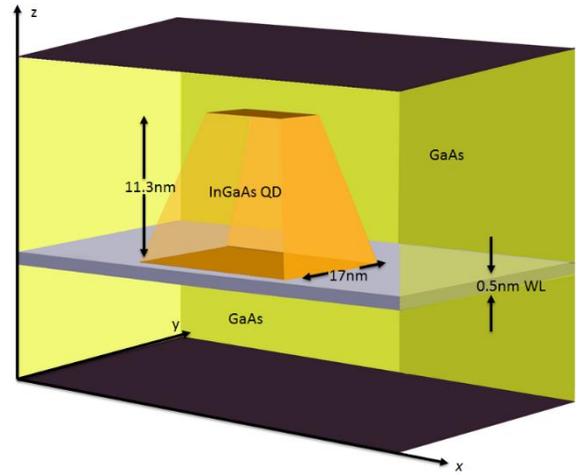

*Figure 1: Profile of a truncated pyramidal $In_xGa_{1-x}As$ QD with square base of the area $17nm \times 17nm$ and the height of 2/3 times the base width and $0.5nm$ wetting layer.*

The parameters for bulk materials which are used in this paper are presented in Table 1 [34-36].

| Parameter | GaAs | InAs |
|---|---|---|
|  |  |  |
| Effective electron mass ($\Gamma$) | 0.067m$_o$ | 0.026m$_o$ |
| Effective heavy hole mass | 0.5m$_o$ | 0.41m$_o$ |
| lattice constant | 0.565325 nm | 0.60583 nm |
| Elastic constants | $C_{11} = 122.1$ $C_{12} = 56.6$ $C_{44} = 60$ | $C_{11} = 83.29$ $C_{12} = 45.26$ $C_{44} = 39.59$ |
| Band gap (0K) | 1.424eV | 0.417eV |

*Table 1: Parameters used in the model*

For different compositions of indium in $In_xGa_{1-x}As$, the effective masses are calculated at 27°C as follow:

Effective electron mass is [37]

$$m_e = (0.023 + 0.037x' + 0.003x'^2)m_o \quad (12)$$

Effective heavy-hole mass is [38]

$$m_h = (0.41 + 0.1x')m_o \quad (13)$$

Also, the lattice constant is [25, 39]:

$$a = (0.60583 - 0.0405x') \, nm \quad (14)$$

where $x' = 1 - x$.

## III. RESULTS AND DISCUSSION ON EFFECT OF INDIUM PERCEBTAGE

Both GaAs and InAs have direct band-gaps. Figures 2(a,b,c,d) illustrate the band edge at all points of the pyramidal QDs at 27°C for conduction band edge $\Gamma$ and valence Heavy-Hole (HH) band when indium percentage changes from 10% to 100%. Obviously, the band edges change in different points. In figures 2(a,b) the white colors around the QD inform of the existence of an obstacle for electrons in the conduction band (CB). However, increase of indium composition from 10% to 100% has led to lower value of $\Gamma$ band edge. The reverse trend is seen by comparison of figures 2(c,d); as it is clear increase of indium content gives rise in the higher edge of the valence band (VB). Band edge of the surrounding GaAs at all figures has been also subjected to change at the points closer to the QD, which is specially approved near vertices.

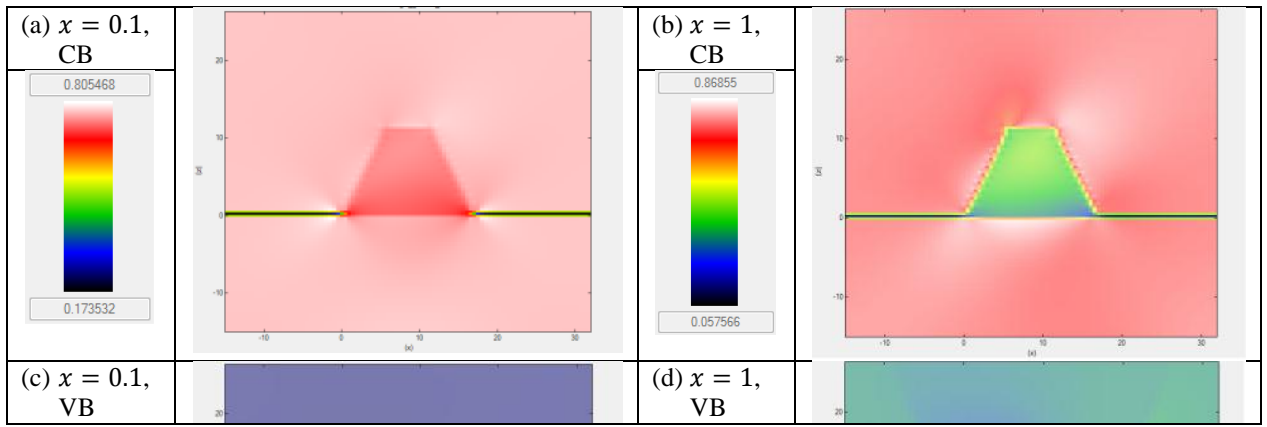

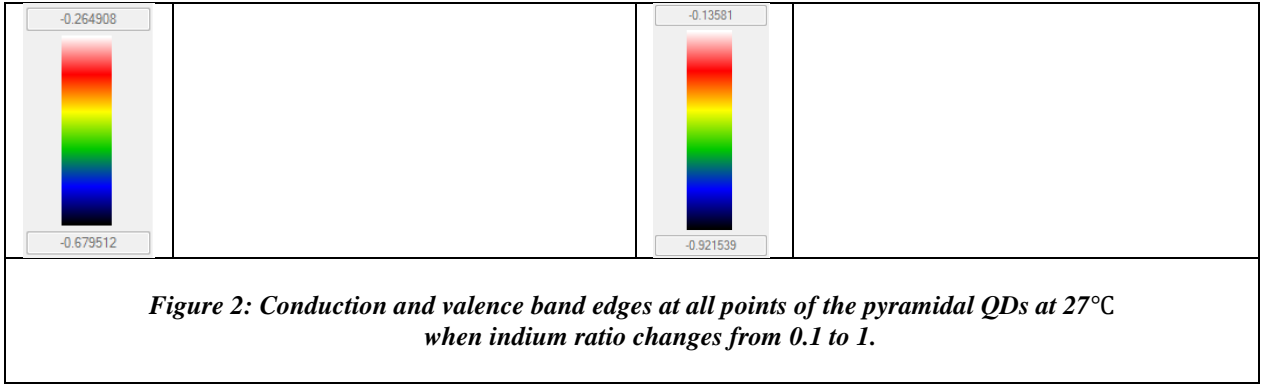

Figure 2: *Conduction and valence band edges at all points of the pyramidal QDs at 27°C when indium ratio changes from 0.1 to 1.*

In figure 3 conduction and valence band edges are shown along z-direction together with the first allowed energy state for electrons and holes. By inclusion of more indium, as it is seen in figure 3(a,b,c,d), more separation occurs between the band edges related to the InGaAs QD and the surrounding GaAs. In addition, all band edges approach to each other at higher values of indium percentage.

More interestingly, by rising the indium content to $x = 0.3$ (figure 3(b)), the first electron state lays into the quantum dot, and it go deeper into the QD at $x = 0.7$ (figure 3(c)); for $x = 1.0$ energy of the ground state lowers more. This ground state (GS) energy can be used in the recombination energy of electron-holes (e-h). The e-h transition energy (Δ) has been also subjected to change by indium ratio. This leads to different wavelengths of the QD laser. In the same way, it can be observed that all the band edges and the band-gap have been sensitive to indium percentage.

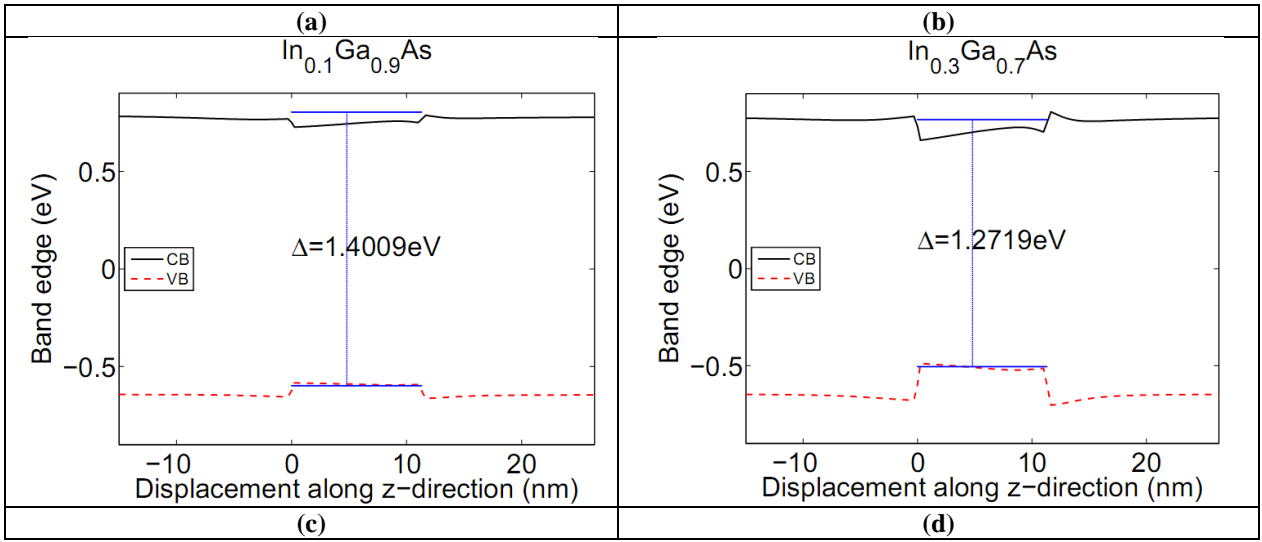
5

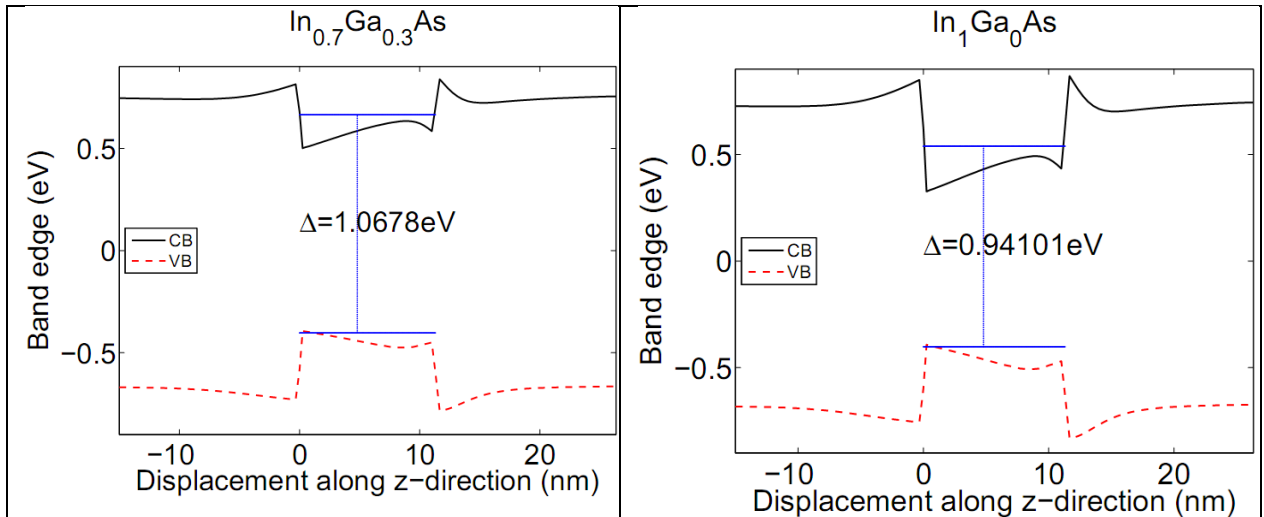

Figure 3: Conduction and valence band edges together with the first eigenvalue for electrons and holes.

Moreover, variation of conduction and valence band edges are shown respectively in figures 4(a) and 4(b). Clearly, indium increase has resulted in lower values of electronic band edge and enhanced value of heavy-hole band edges. It is also notable that effect of indium percentage on the heavy-hole band edge is not the same for $x$ values of more than ~50% and the one relating x<~50%. However, indium inclusion sounds to **linearly** decrease the conduction band edge.

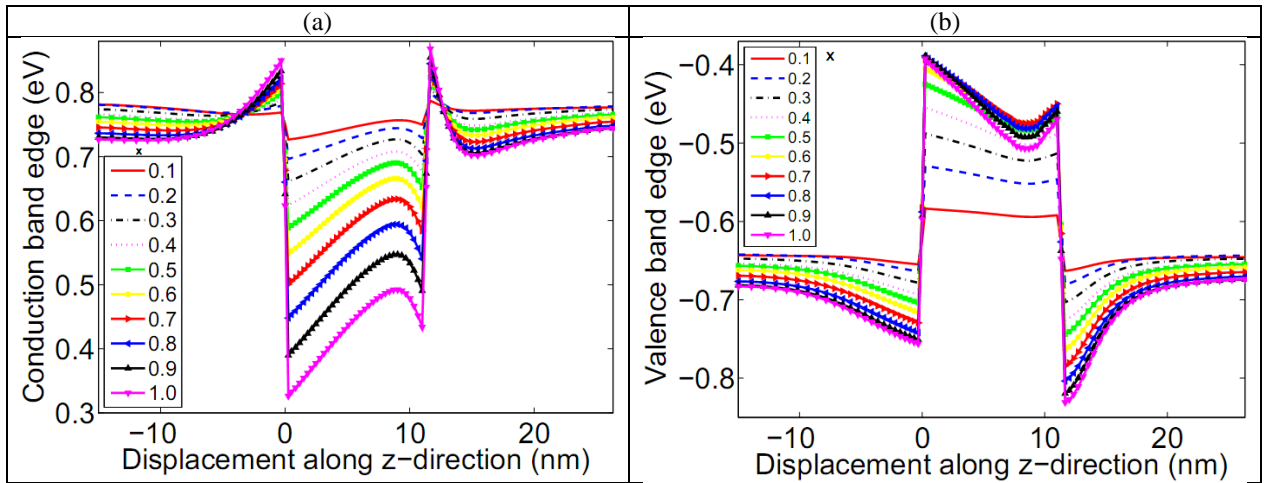

Figure 4: Conduction and valence band edges for various amounts of indium ratio.

For a better estimation of the behavior of band edges and e-h states in terms of indium ratio, valence and conduction bands together with e-h energies are plotted in figure 5(a) as a function of indium percentage. In addition, the energy gap and recombination energy of the first eigenvalue are drawn as a function of indium content in figure 5(b). The most interesting result of this work is that the effect of indium percentage on recombination energy and the band gap is not linear. Decrease of recombination energy when composing more indium results in the elongated laser wavelength which must be paid attention in the lasers. The other result which is clear from figure 5(b) is that decrease of the band gap energy is not similar to that of e-h recombination energies, since the slope of its curves is larger than recombination energies.

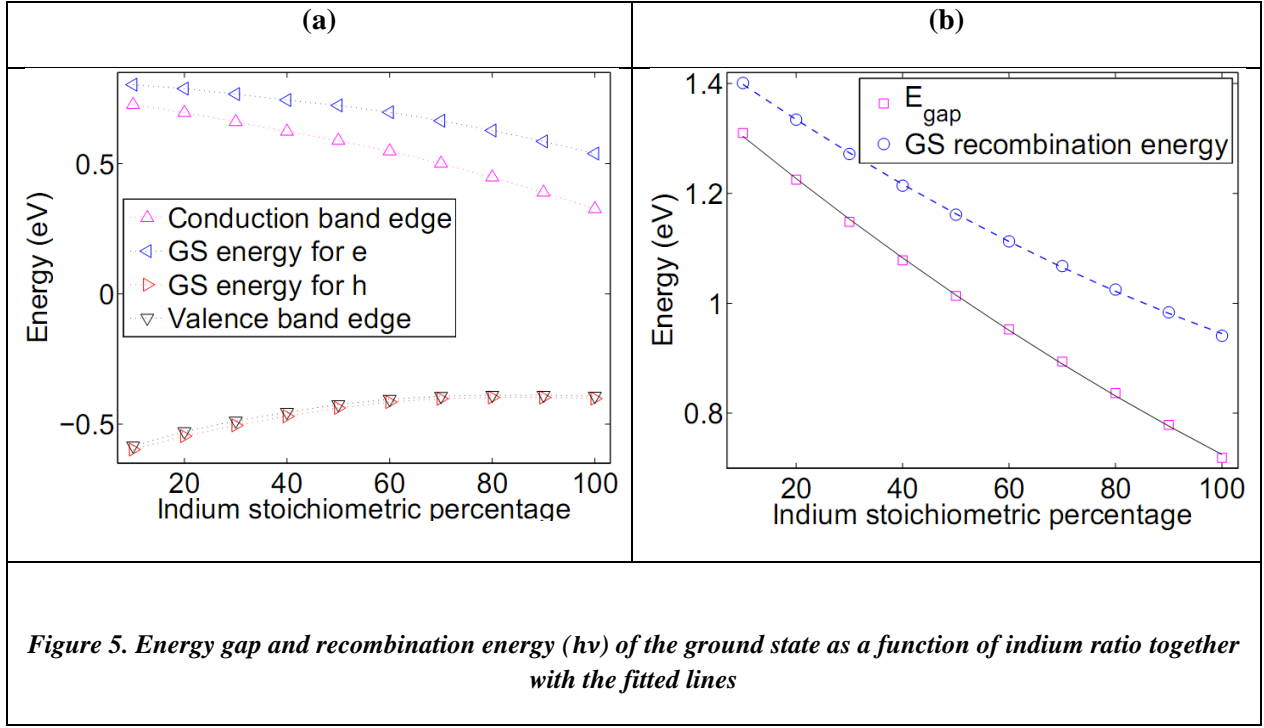

*Figure 5. Energy gap and recombination energy (hν) of the ground state as a function of indium ratio together with the fitted lines*

For different ratios of indium in $In_xGa_{1-x}As$, the behavior of energy gap for bulk materials is [40]:

$$E_g(x) = aX^2 - bX + c \quad at\ 27°C \quad (15)$$

in which $X$ is the percentage of indium content. Coefficients $a, b,$ and $c$ are represented in Table 2 for comparison with our results for a nano-scale truncated pyramid obtained from the lines fitted to our data in figure 7(b).

| Coefficient | $E_g$ for bulk material | $E_g$ in our QD | e-h recombination energy |
|---|---|---|---|
| $a$ | 0.000043 | 0.0000157 | 0.00001697 |
| $b$ | 0.0149 | 0.008162 | 0.006907 |
| $c$ | 1.42 | 1.384 | 1.466 |

*Table 2: Coefficients of equation 4 for bulk material and for our results*

Comparison of our results with this formula shows that our results can be logical, since on the one hand the lowering effect of indium is clear from this equation, and on the other hand, the nonlinear term is appeared here too.

Strain is a key factor in optimizing the optical and electronic properties of semiconductor nanostructures. By taking into account the effect of length variation along all directions, components of the strain tensor read:

$$\varepsilon_{ij}(r) = \frac{u'_{ij}(r)+u'_{ji}(r)}{2} \quad (16)$$

in which $u'_{ij}(r) = \frac{du_i(r)}{dr_j}$ with $i,j$ taking values 1, 2, 3 respectively corresponding to $x, y,$ and $z$. $r_i$ is the length along direction $i$; $du_i$ is displacement in direction $i$ due to lattice deformation [41]. Although the distortion matrix $u$ may be non-symmetric, this tensor is real and symmetric (i.e., $\varepsilon_{ji} = \varepsilon_{ij}$). Diagonal components are associated with expansion/compression per unit length along an axis which refers to compressive-strain if negative and tensile-strain if positive; in this case, angles of the volume element remain fixed but the side lengths change. Off-diagonal components are related to rotation and shear deformations which lead to variation of angles while the volume remains fixed. The general strain tensor is a 3 × 3 matrix,

$$\varepsilon = \begin{bmatrix} \varepsilon_{xx} & \varepsilon_{xy} & \varepsilon_{xz} \\ \varepsilon_{xy} & \varepsilon_{yy} & \varepsilon_{yz} \\ \varepsilon_{xz} & \varepsilon_{yz} & \varepsilon_{zz} \end{bmatrix} \quad (17)$$

For substrate index (001) the biaxial strain parallel to interface is [42]:

$$\varepsilon_{xx} = \varepsilon_{yy} = \varepsilon_{\parallel} = \frac{a_{substrate}-a_{QD}}{a_{QD}} \quad (18)$$

in which $a_{substrate}$ and $a_{QD}$ are respectively the substrate and quantum dot lattice constants.

The uniaxial strain component $\varepsilon_{zz}$ perpendicular to the interface is obtained as:

$$\varepsilon_{zz} = \varepsilon_{\perp} = -\frac{2C_{12}}{C_{11}} \varepsilon_{\parallel} \quad (19)$$

which is obtained at zero stress $\sigma_{zz}$. $C_{ij}$ are elements of the matrix of elasticity constants which correlates stress $\sigma$ to strain by Hooke's law (i.e., $\sigma = C\varepsilon$) [43, 44]. This relation for a zinc-blende crystal is given as [45]:

$$\begin{bmatrix} \sigma_{xx} \\ \sigma_{yy} \\ \sigma_{zz} \\ \sigma_{xy} \\ \sigma_{xz} \\ \sigma_{yz} \end{bmatrix} = \begin{bmatrix} C_{11} & C_{12} & C_{12} & 0 & 0 & 0 \\ C_{12} & C_{11} & C_{12} & 0 & 0 & 0 \\ C_{12} & C_{12} & C_{11} & 0 & 0 & 0 \\ 0 & 0 & 0 & 2C_{44} & 0 & 0 \\ 0 & 0 & 0 & 0 & 2C_{44} & 0 \\ 0 & 0 & 0 & 0 & 0 & 2C_{44} \end{bmatrix} \begin{bmatrix} \varepsilon_{xx} \\ \varepsilon_{yy} \\ \varepsilon_{zz} \\ \varepsilon_{xy} \\ \varepsilon_{xz} \\ \varepsilon_{yz} \end{bmatrix} \quad (20)$$

Figures 6(a) and (b) compare the $\varepsilon_{zz}$ elements of the strain tensor for two different values of the indium ratio which indicate the more absolute value of $\varepsilon_{zz}(x,y,z)$ at the interfaces and vertices. Also, by increasing the indium percentage, it is seen that not only the absolute value of strain increases but also its impact extends more into the points far from interfaces. In addition, figure 6(c) and (d) illustrates $\varepsilon_{xx}$ for low indium and full InAs/GaAs. Equation (5) denotes that negative (positive) value of strain tensor components represent compressive (tensile) strain [41]. As it is observable, strain is compressive inside and tensile outside the QD.

Strain is due to the two different types of materials connected to each other at the interface. On the one hand, strain depends on lattice constants of neighbor materials; on the other hand, lattice constant of indium is larger than arsenide and they both are larger than gallium. So, increase of indium percentage is expected to increase the mean lattice constant and so the strain.

| a) $\varepsilon_{zz}$ for $x = 0.1$ | | c) $\varepsilon_{xx}$ for $x = 0.1$ |
|---|---|---|

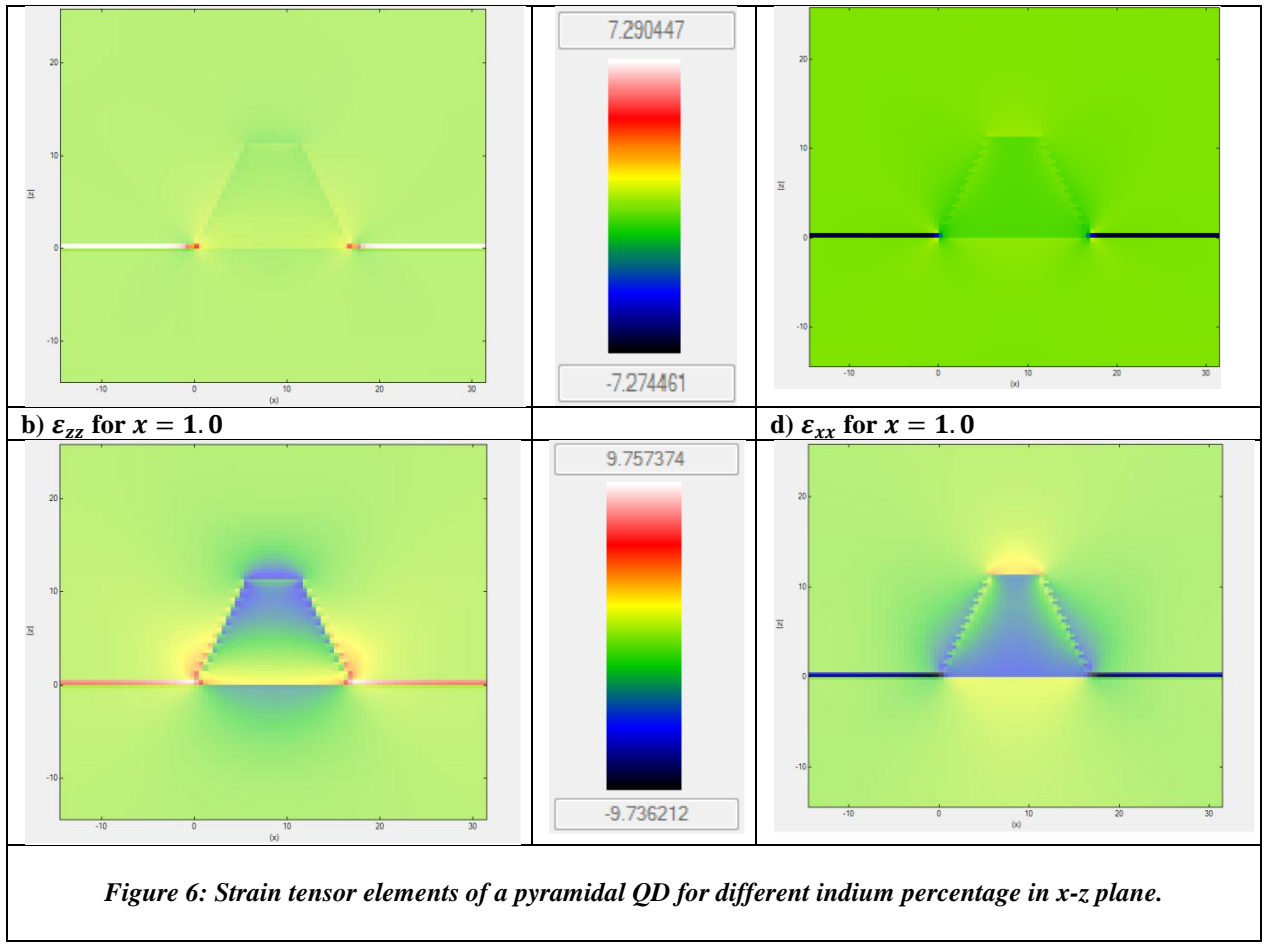

*Figure 6: Strain tensor elements of a pyramidal QD for different indium percentage in x-z plane.*

Figures 7(a,b) show the non-zero elements of the strain tensor along z-direction at the middle of the structure. It can be noticed that the mismatch due to existence of indium in one side of interfaces has lead to jump in the strain tensor for $\varepsilon_{xx}$ ($\varepsilon_{yy}$) and $\varepsilon_{zz}$, which shows a tension in GaAs and compression in QD lattice constants. This figure, as clear is sensitive to the indium composition. By indium increase, the absolute values of the strain components have generally grown, showing the effect of increase in the mismatch made due to the more percentage of indium with a lattice constant of more than gallium. This result seems logical, since as it was proved, strain is dependent to the lattice constants which linearly increase by indium inclusion (see equation 3). Therefore this can be acceptable in the previous analytical pints of view.

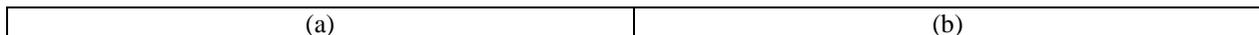

| (a) | (b) |

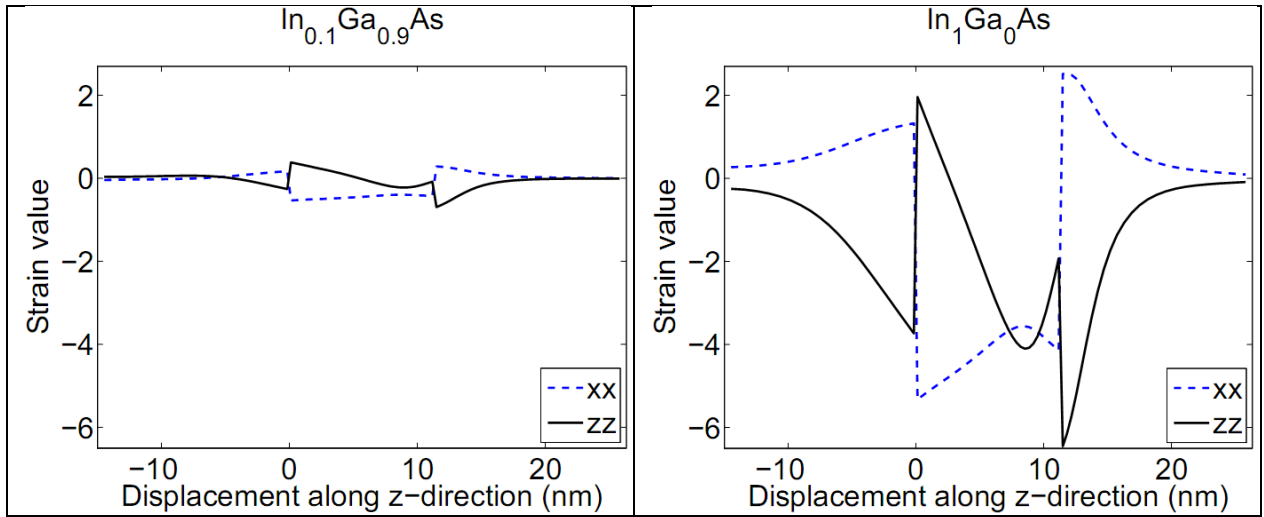

Figure 7: Nonzero elements of the strain tensor at (a) $x = 0.1$ and (b) $x = 1.0$.

Also, figure 8 depicts the strain tensor components along x-axis for different values of indium content. Clearly, along x-direction the strain components $\varepsilon_{xx}$ and $\varepsilon_{yy}$ are not the same at all points any more. As it is found it was true only at the middle points of the QD. In addition, component $\varepsilon_{xz}$ does not vanish along x-axis. As mentioned, this component represents a rotational strain which has its maxima at the interface. The tensile strain component $\varepsilon_{zz}$ along $x - axis$ is viewed.

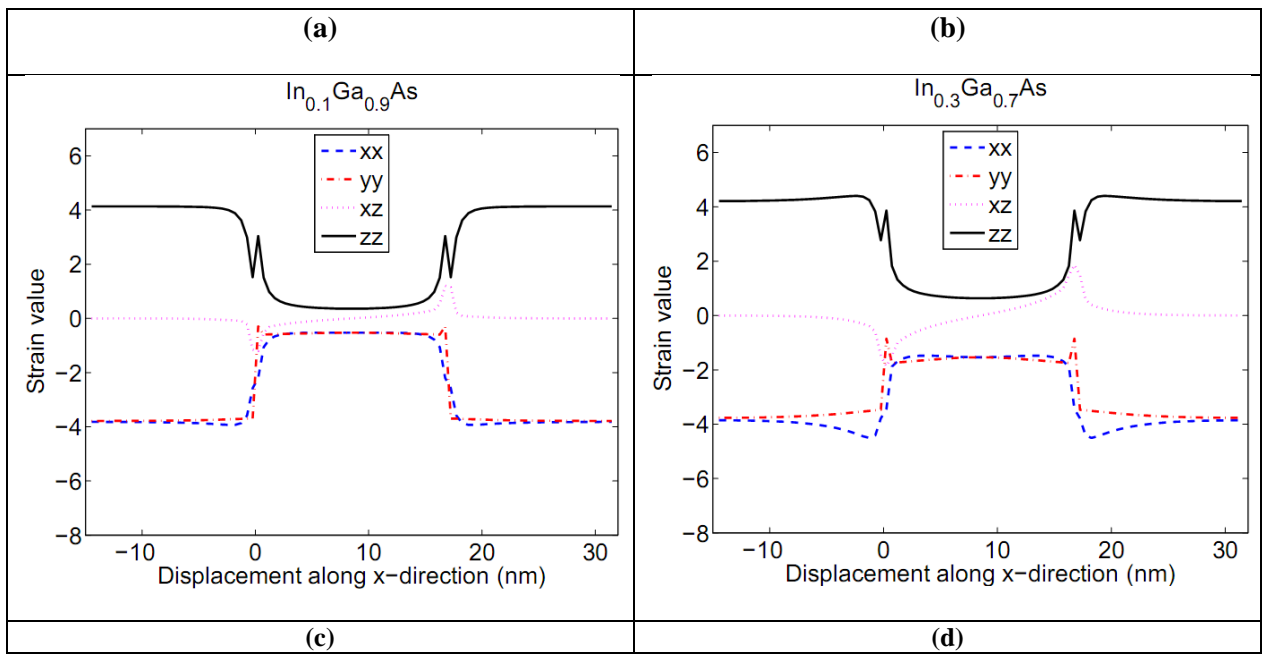

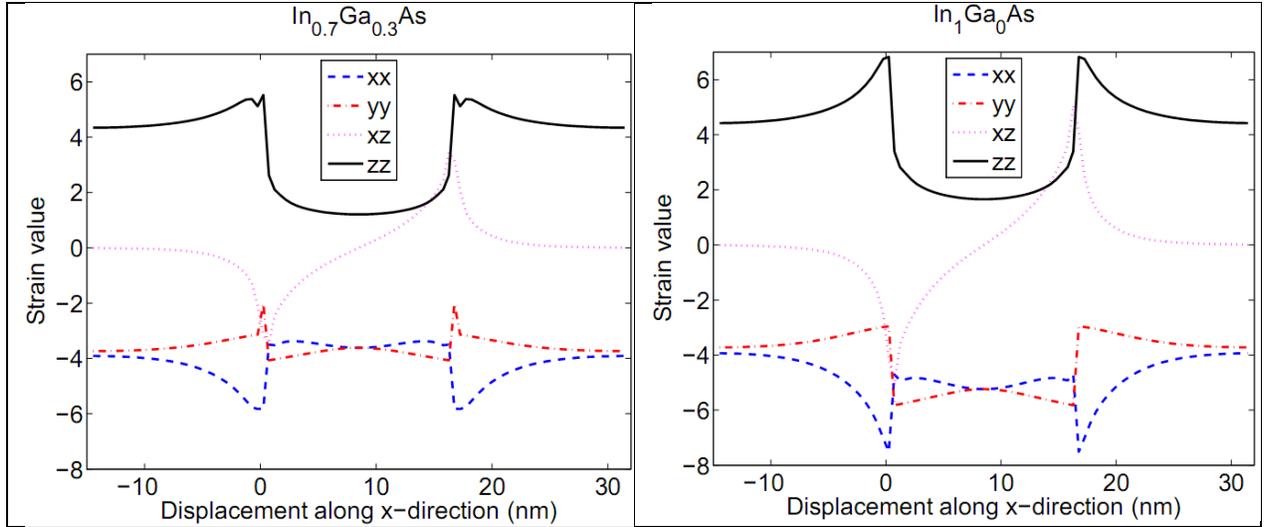

*Figure 8: Nonzero elements of the strain tensor at (a) $x = 0.1$, (b) $x = 0.3$, (c) $x = 0.7$, and (d) $x = 1.0$.*

## IV. CONCLUSION

We surveyed the band structure and strain tensor of $In_xGa_{1-x}As$ QDs grown on GaAs substrate by quantum numerical solution. It was proved that increase of indium composition lowers the band gap and of electron-hole recombination energy. The main result was that the decline is nonlinear and with different slope of the band gap and e-h recombination energy. Moreover, it was shown that strain tensor is diagonal along z, and the absolute value of the components becomes larger by more indium inclusion. Our results appear to be in very good consonance with similar researches.

## ACKNOWLEDGEMENT

The authors give the sincere appreciation to Dr. S. Birner for providing the advanced 3D Nextnano++ simulation program [30] and his instructive guides. We would like to acknowledge many useful discussions with Prof. S. Farjami Shayesteh, K. Kayhani, and Y. Yekta Kia.